\documentclass{article} \usepackage{fortschritte} \usepackage{amsmath}
\usepackage{amssymb}
\usepackage{graphics}
 \def\C{{ \mathbb C}} \def\R{{ \mathbb R}}     
    \def\tr{\operatorname{tr}}

 \def\tr{{\rm tr\, }}

\def\id{\protect{{1 \kern-.28em {\rm l}}}}

\newcommand{\be}{\begin{equation}} \newcommand{\ee}{\end{equation}}
\newcommand{\bea}{\begin{eqnarray}} \newcommand{\eea}{\end{eqnarray}}
\newcommand{\beann}{\begin{eqnarray*}}
  \newcommand{\eeann}{\end{eqnarray*}}
\newcommand{\bfig}{\begin{figure}} \newcommand{\efig}{\end{figure}}

\newcommand{\ba}{\begin{array}}\newcommand{\ea}{\end{array}}

\newtheorem{Proposition}{Proposition}[section]

\newtheorem{Theorem}{Theorem}[section]
\newtheorem{Lemma}{Lemma}[section]
\newtheorem{Corrolary}{Corrolary}[section]

\newcommand{\bp}{\begin{Proposition}}
  \newcommand{\ep}{\end{Proposition}}
\newcommand{\bt}{\begin{Theorem}} \newcommand{\et}{\end{Theorem}}
\newcommand{\bl}{\begin{Lemma}} \newcommand{\el}{\end{Lemma}}
\newcommand{\bc}{\begin{Corrolary}} \newcommand{\ec}{\end{Corrolary}}

   \def\ep{\eps}
 \def\hn{{ \hat N}}     
\def\hx{\hat X}\def\hy{\hat Y}\def\hf{\hat f}
\def\cW{{\cal W}}

\def\erf#1{(\ref{#1})}

\begin {document}                 
\def\titleline{
%%%%%%%%%%%%%%%%%%%%%%%%%%%%%%%%%%%%%%%%%%%%%%
%                                            
% Insert now the text of your title.         
% Make a linebreak in the title with         
%                                            
%            \newtitleline                   
%                                            
%%%%%%%%%%%%%%%%%%%%%%%%%%%%%%%%%%%%%%%%%%%%%%%
  Konishi Anomalies and $N=1$ Curves
%%%%%%%%%%%%%%%%%%%%%%%%%%%%%%%%%%%%%%%%%%%%%%% 
}
\def\email_speaker{ {\tt
%%%%%%%%%%%%%%%%%%%%%%%%%%%%%%%%%%%%%%%%%%%%%%
%                                                  
% Insert now the e-mail address of the speaker or  
% the author that should get the electronic mail   
% of the publishing house                           
%                                                  
%%%%%%%%%%%%%%%%%%%%%%%%%%%%%%%%%%%%%%%%%%%%%%         
    Karl.Landsteiner@uam.es
%%%%%%%%%%%%%%%%%%%%%%%%%%%%%%%%%%%%%%%%%%%%%%         
  }} \def\authors{
%                                            
%  Insert now the name (names) of the author 
%  (authors).                                
%  In the case of several authors with       
%  different addresses use labels e.g.       
%                                            
%             \1ad  , \2ad  etc.             
%                                            
%  to indicate that a given author has the   
%  address number 1 , 2 , etc.               
%  Signify the person with the above e-mail  
%  address with \sp (behind the address      
%  labels, e.g.                              
%                                            
%            \2ad\sp                         
%                                            
%  when the 2nd author                       
%  happens to be the 'speaker')                                  
%%%%%%%%%%%%%%%%%%%%%%%%%%%%%%%%%%%%%%%%%%%%%%         
  K. Landsteiner\sp
%%%%%%%%%%%%%%%%%%%%%%%%%%%%%%%%%%%%%%%%%%%%%%         
} \def\addresses{
%%%%%%%%%%%%%%%%%%%%%%%%%%%%%%%%%%%%%%%%%%%%%%
%                                            
% List now the address. In the case of       
% several addresses list them by numbers     
% using e.g.                                 
%                                            
%             \1ad , \2ad   etc.             
%                                            
% to numerate address 1 , 2 , etc.           
%                                            
%%%%%%%%%%%%%%%%%%%%%%%%%%%%%%%%%%%%%%%%%%%%%
  Instituto de F\'\i sica Te\'orica, Universidad Aut\'onoma de Madrid \\
  28049 Madrid, Spain \\
%%%%%%%%%%%%%%%%%%%%%%%%%%%%%%%%%%%%%%%%%%%%%
} \def\abstracttext{ We present a brief summary of exact results on
  the non-perturbative effective superpotential of $N=1$
  supersymmetric gauge theories based on generalized Konishi anomaly
  equations. In particular we consider theories with classical gauge
  groups and chiral matter in two-index tensor representations. All
  these theories can be embedded into theories with unitary gauge
  group and adjoint matter. This embedding can be used to derive
  expressions for the exact non-perturbative superpotential in terms
  of the $1/N$ expansion of the free energy of the related matrix
  models.  }

%%%%%%%%%%%%%%%%%%%%%%%%%%%%%%%%%%%%%%%%%%%%%%%%%%%%%%%%%%%%%%
%
% Take out Preprint numbers before submitting to Proceedings
%
%
%\vspace*{-2truecm}
%\begin{flushright} \tt
  IFT-UAM/CSIC-04-63
%\end{flushright}
%%%%%%%%%%%%%%%%%%%%%%%%%%%%%%%%%%%%%%%%%%%%%%%%%%%%%%%%%%%%%%
%%%%%%%%%%%%%%%%%%%%%%%%%%%%%%%%%%%%%%%%%%%%%%%%%%%%%%%%%%%%%%

\large \makefront
%%%%%%%%%%%%%%%%%%%%%%%%%%%%%%%%%%%%%%%%%%%%%%%%
%                                              %
%  Insert now the remaining parts of           %
%  your article.                               %
%                                              %
%%%%%%%%%%%%%%%%%%%%%%%%%%%%%%%%%%%%%%%%%%%%%%%%
\section{Introduction}

Dijkgraaf and Vafa conjectured in \cite{DV} that the exact
non-perturbative superpotential of a $N=1$ supersymmetric gauge theory
can be computed in terms of the free energy of a related matrix model.
The original example was based on a theory with $SU(N)$ gauge group
and matter in the adjoint representation.  In \cite{cdsw} it was
pointed out that the loop equations in the large $N$ limit of the
matrix model are equivalent to a set of generalized Konishi anomaly
relations truncated to the chiral ring of the theory in question (see also
\cite{Gorsky}).
This was extended to theories with classical gauge groups and chiral
matter multiplets in more general representations.  A particular
interesting example proved to be the theory based on symplectic gauge
groups $Sp(N)$ and matter transforming in the antisymmetric
representation \cite{alday}. These theories presented a puzzle as pointed out in
\cite{kraus}. The matrix model free energy did not seem to reproduce
the effective superpotential in the cases that were already known from
field theory arguments based on holomorphy. This puzzle could be
resolved by Cachazo in \cite{cachazo} using an embedding of the theory
with symplectic gauge group and antisymmetric matter into a theory
with unitary gauge group and adjoint matter. This embedding is most
naturally understood in the string realization of the theories and the
earlier noted discrepancy could be traced back to nontrivial
contributions of orientifolds \cite{ll}, \cite{ikrsv}. The string
theory realization also allowed a better understanding of the
embedding and an extension to all theories with orthogonal or
symplectic gauge group and two-index matter.

What was missing so far in the literature was the case of unitary
gauge groups with symmetric or antisymmetric matter. It has been
treated now in \cite{adi}, \cite{land}, \cite{argurio}. Although the
model has a somewhat different flavour from the previously described
ones (since it needs two chiral multiplets in conjugate
representation) it can be treated in a completely analogous way by
using a composite adjoint field.

\section{Konishi Anomalies}
\label{konishis}

We will now briefly summarize the main features of the generalized
Konishi anomaly equations of the theories with classical gauge groups
and a single matter multiplet in a two-index tensor representation.

The theories under consideration have a superpotential of the form
\begin{equation}
  \label{eq:superpot}
  W = \sum_{i}^{d} \frac{g_k}{k+1} \tr ( \Phi^{k+1} )
\end{equation}
where $\Phi$ is a chiral superfield transforming as $\Phi \rightarrow
g \Phi g^{-1}$ under the a gauge transformation $g$. For unitary ague
groups that means it transforms under the adjoint and for symplectic
and orthogonal it should be a symmetric or antisymmetric two-index
tensor representation.

Two generating functions of correlators of chiral operators are of
particular interest. They are defined as
\begin{equation}
  \label{eq:ftresolvents}
T(z) = \left\langle\tr \left( \frac{1}{z-\Phi} \right)\right\rangle ~~~,~~~ 
R(z) = -\frac{1}{32\pi^2} \left\langle\tr 
\left( \frac{\cW^\alpha \cW_\alpha}{z-\Phi} \right)\right\rangle\,.
\end{equation}
where $\cW^\alpha$ is the gaugino.  Note that the first $d+2$
coefficients of the Laurent expansion of $T(z)$ are the operators
defining the superpotential. The strategy is therefore to determine
the generating functions $T(z)$ and $R(z)$. Then we can integrate the
differential equations $\frac{\partial W_{eff}}{\partial g_k} = \tr
(\Phi^{k+1})$ and determine the effective superpotential up to a term
that is independent of the couplings $g_k$.

Since we need to know only correlators of chiral operators we can
restrict ourselves to the chiral ring. The anomaly equation which
allows the determination of $T(z)$ and $R(z)$ gives rise to a relation
in the chiral ring
\begin{equation}
  \label{eq:anomaly}
\left\langle
\delta \Phi_I \frac{\partial W}{\partial \Phi_I}\right\rangle = 
- \frac{1}{32\pi^2} 
\left\langle\cW^\alpha_I\,^J\cW_{\alpha,J}\,^K \frac{\partial( 
\delta \Phi_K)}{\partial \Phi_I}\right\rangle   
\end{equation}
The left hand side is the response of the classical theory to a
arbitrary holomorphic field transformation $\delta \Phi$ whereas the
right hand side is a (anomalous) quantum effect. The capital indices
denote a basis in the representation of the field $\Phi$. The anomaly
is a generalization \cite{cdsw}, \cite{Gorsky} of the well-known
Konishi anomaly \cite{CPSK}. The anomaly equation \erf{eq:anomaly}
has been shown to be exact non-perturbatively in \cite{svrcek}.

Suitably chosen variations $\delta \Phi$ lead to two relations
\begin{eqnarray}
  \label{eq:general}
    \frac{1}{2} R^2(z) &=& W'(z) R(z) + f(z) \\
     T(z) R(z) &=& W'(z) T(z) + X(R) + c(z)\\ 
\end{eqnarray}
$R(z)$ takes values on a hyperelliptic Riemann surface $\Sigma$ and
$T(z)$ is a meromorphic differential on $\Sigma$, $f(z)$ and $c(z)$
are polynomials
\begin{equation}
  \label{eq:polynomialsfc}
    f(z) = -\frac{1}{32\pi^2} \frac{(W'(z)-W'(\Phi))\cW^2}{z-\Phi} ~~~,~~~
  c(z) = \frac{(W'(z)-W'(\Phi))}{z-\Phi}\,.
\end{equation}
There is a one-to-one correspondence between the coefficients of these
polynomials and the period integrals
\begin{equation}
  \label{eq:periods}
  S_i = \oint_{C_i} R(z) dz 
\leftrightarrow f(z)\, ~~~,~~~
  N_i = \oint_{C_i} T(z) dz  \leftrightarrow c(z) \, .
\end{equation}
The physical interpretation is that the cuts on the Riemann surface
represent a vacuum in which the gauge group $G$ is broken to a
subgroup $\prod_i G_i$.  The rank of each factor group $G_i$ is
determined by $N_i$\footnote{It is $N_i$if the factor group is unitary
  and $N_i/2$ if it is symplectic or orthogonal.} and the gaugino
condensate in a factor group is given by $S_i$ \cite{cdsw}.

The form of the anomaly relations depend on the gauge group and matter
content only through the term labelled as $X(R)$, specifically it is
given by
\begin{itemize}
\item $ U(N)$ and $\phi$ adjoint: $X(R) =0$
\item $SO(N)/SP(N)$ and $\phi$ adjoint: $X(R) = -2 \frac{\epsilon} {z}
  R(z)$, $\epsilon = \pm 1$
\item $SO(N)/SP(N)$ and $\phi^T = \epsilon \phi$: $X(R) = 2 \epsilon
  R'(z)$, $\epsilon = \pm 1$
\end{itemize}
where in the last two lines $\epsilon = +1 $ for $SO(N)$ and $\epsilon
= -1 $ for $SP(N)$. $X(R)$ determines subleading $1/N$ corrections in
the effective superpotential.

Let us now have a look to the new model with symmetric and
antisymmetric tensors of $U(N)$. We take two matrix valued superfields
$X$ and $Y$ obeying
\begin{equation}
  \label{eq:XY}
  X^T = \epsilon X ~~~,~~~ Y^T = \epsilon Y \\
\end{equation}
and transforming as $X\rightarrow U X U^T$ and $y \rightarrow
U^* Y U^\dagger$ under $U(N)$ gauge transformation.  We can form the
composite adjoint $\Phi = XY$ and define a superpotential and
 the generating functions of
chiral correlators just as in \erf{eq:superpot}, \erf{eq:ftresolvents}.  To discuss the
classical theory it is useful to compute the F-flatness conditions as
$X\frac{\partial}{\partial X} W(\Phi) = \Phi W'(\Phi)$.  This shows
that the eigenvalues of the composite adjoint have either to vanish or
be one of the $d$ roots of $W'(z)=0$. The classical gauge symmetry
breaking pattern is

\begin{eqnarray}
  \label{eq:pattern}
  U(N) \rightarrow \left\{  
\begin{array}{cc}
U(N_0) \bigotimes_{i=1}^d SO(N_i) , & \epsilon = +1 \\
U(N_0) \bigotimes_{i=1}^d SP(N_i) , & \epsilon = -1 \\
\end{array}\right.
\end{eqnarray}
The unitary factor group corresponds to the $N_0$ fold degeneracy of
the zero-eigenvalue of $\Phi$. Since the non-zero eigenvalues
ultimately come from symmetric or antisymmetric matrices the gauge
group is broken to factors of either orthogonal or symplectic groups.

The variations
\begin{equation}
  \label{eq:variations}
  \delta X = \frac{1}{z-XY}X  ~~~,~~~
 \delta X = -\frac{1}{32\pi^2}\frac{\cW^2}{z-XY}X 
\end{equation}
lead to the anomaly relations
\begin{eqnarray}
  \label{eq:konishis_XY}
   \frac{1}{2} z R^2(z) &=& zW'(z) R(z) +  f(z) \\
    zT(z) R(z) &=& zW'(z) T(z) + 2\epsilon z R'(z) + c(z)
\end{eqnarray}
where the polynomials $f(z)$ and $c(z)$ are defined by
\begin{equation}
  \label{eq:poly_XY}
 f(z) = -\frac{1}{32\pi^2} \frac{(zW'(z)-XYW'(XY))\cW^2}{z-XY}~~,~~
  c(z) = \frac{(zW'(z)-XYW'(XY))}{z-XY}
\end{equation}
We also note that the Riemann surface $\Sigma$ defined by
\erf{eq:konishis_XY} has a fixed branch point at $z=0$.

\section{The Matrix Model}
\label{matrixmodel}
The partition function of the related matrix model is given by
\begin{equation}
  \label{eq:mmi}
  Z = \frac{1}{|G|} \int_\Gamma d\hx d\hy e^{-\frac{1}{\kappa} 
\tr[W(\hx\hy)]}
\,,
\end{equation}
where $\hx$ and $\hy$ are the matrices corresponding to the chiral
multiplets $X,Y$ of the gauge theory. They are thus complex
$\hn\times\hn$ matrices obeying $(\hx^T, \hy^T) = \epsilon (\hx,
\hy)$.  $|G|$ is a normalization factor including the volume of the
gauge group and $\Gamma$ is a suitably chosen path in the
configuration space ${\cal M}$ of the matrices $\hx, \hy$ with
$\mathrm{dim}_{\R}(\Gamma) = \mathrm{dim}_{\C}({\cal M}) $.

We define the matrix model resolvent $\omega(z) =
\kappa\tr(\frac{1}{z-\Phi})$ and the polynomial $\hat f (z) = - \kappa
\tr( \frac{ z W'(z) - \Phi W'(\Phi)}{z-\Phi} )$. Standard arguments
\cite{land} lead then to the exact loop equation
\begin{equation}
  \label{eq:loop}
  \left\langle
     \frac{1}{2} z \omega^2(z) -zW'(z) \omega(z) -
     \frac{\epsilon}{2} z \omega'(z)-\hf(z)\right\rangle =0
\end{equation}

The large $\hn$ (or equivalently $\kappa$) expansion $\langle
\omega(z)\rangle = \sum_{k=0}^\infty \kappa^k \omega_k(z)$
%of the loop equation gives up to $O(\kappa)$:
%\begin{eqnarray}
%\label{eq:expansio}
%\frac{z}{2} \omega_0(z)^2 - z W'(z) \omega_0(z) - \hf(z) &=& 0 ~~,~~ O(0)\\
%\omega_0(z)\omega_1(z)-\frac{\epsilon}{2}z\omega'_0(z) - z W'(z) \omega_1(z) &=&
%0 ~~,~~ O(\kappa)
%\end{eqnarray}
%Consider also the operator $\delta = \sum_i N_i \frac{\partial}{\partial S_i}$
%and apply to the equation for $\omega_0(z)$,
%\begin{equation}
%  \label{eq:delta_omega}
%  z\omega_0(z) \delta\omega_0(z) - z W'(z)\delta\omega_0(z)  - \delta \hf(z) = 0
%\end{equation}
%We can then see that we recover the anomaly relations of the gauge theory
%if we identify:
leads to an identification of the matrix model quantities with the
field theory quantities according to
   \begin{eqnarray}\label{eq:identify}
     \omega_0(z) = R(z) &,& \delta\omega_0(z) + 4 \omega_1(z) = T(z) \\
     \hf(z) = f(z) &,& \delta \hf(z) = c(z) \\
     \mathrm{filling~fraction}~\kappa \hn_i &=& 
     \mathrm{gaugino~condensate}~ S_i
   \end{eqnarray}
   where we used the differential operator $\delta = \sum_{i=0}^{d}
   N_i \frac{\partial} {\partial S_i}$.  Finally the superpotential of
   the field theory can be computed from the matrix model free energy
   as
\begin{equation}
  \label{eq:weff}
  W_{eff} = \sum_{i=0}^d N_i \frac{\partial F_0}{\partial S_i} + 4 F_1 + \tau S
\end{equation}
where $F_{0,1}$ are the leading and subleading contributions to the
matrix model free energy in the $1/\hn$ expansion and $\tau$ is the
tree level gauge coupling \cite{DV}.

\section{Superpotential}
\label{superpotentials}
 
In order to give a prescription of how to compute the superpotential
we will follow \cite{cachazo}, \cite{ll}, \cite{ikrsv} and embed the
theory into one with unitary gauge group and elementary adjoint
matter. It is useful now to go to a double cover of the $z$-plane and
introduce
\begin{equation}\label{eq:double}
 \zeta^2 = z~~,~~
  \bar R(\zeta) = \zeta R(\zeta^2)~~,~~  \bar T (\zeta) = \zeta T(\zeta^2)~~,~~
 \bar W(\zeta) = \frac 1 2 W(\zeta^2)
\end{equation}
Except for the cut touching the origin $z=0$ this doubles the number
of cuts as indicated in the following figure.

\begin{figure}[htbp]
  \centering \scalebox{.5}{\input{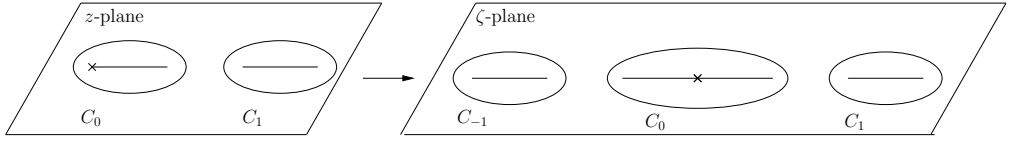}}
  \caption{Going to the double cover}
  \label{fig:1}
\end{figure}

The period integrals are
\begin{eqnarray}
  \label{eq:periodsdouble}
  \oint_{C_0} \bar R(\zeta) d\zeta = S_0 &,& \oint_{C_0} \bar T(\zeta) d\zeta = N_0 \\
\oint_{C_i} \bar R(\zeta) d\zeta = \frac{S_i}{2} &,&
\oint_{C_i} \bar T(\zeta) d\zeta = \frac{N_i}{2} 
\end{eqnarray}

We introduce now $\bar y^2 = (\bar W')^2 + \bar f$, $\bar T = \tilde T
+ \Psi$ and $\tilde c = \bar c + \bar W''-\frac{\bar W'}{\zeta}$.
where
\begin{equation}
  \label{eq:psi}
  \tilde T = \frac{\tilde c}{\bar y} ~~,~~\Psi = \epsilon
(\frac{\bar y'}{\bar y}-\frac{1}{\zeta})
\end{equation}
The reason for introducing this new variables is that the pair $(\bar
y, \tilde T )$ fulfill now the {\em same relations} as the pair $(y =
R + W', T)$ of a theory with unitary gauge group and elementary
adjoint matter! We have therefore a rather specific embedding of the
original theory into one with unitary gauge group and elementary,
adjoint matter. Notice that the period integrals of $\tilde T$ and
$\bar T$ are related by
\begin{equation}
  \label{eq:shift}
  \oint \bar T = \oint \tilde T +\oint \Psi
\end{equation}
and the ranks of the product gauge groups are then determined by
$$
\left.
\begin{array}{c}
N_0\\ \frac{N_i}{2} 
\end{array}\right\}
= \tilde N_i + \epsilon(1-\delta_{i,0})
$$
The superpotential of embedding theory can be computed in the
well-known way \cite{DV} whereas the original theory has an
additional contribution from $1/N$ corrections due to non-orientable
graphs. The relevant terms are
\begin{eqnarray}\label{eq:superpots}
W_{eff}^F &=& \sum_{i=0} \tilde N_i \frac{\partial F_0}{\partial \tilde S_i} =
N_0  \frac{\partial F_0}{\partial \tilde S_0}  +\sum_{i>0} 2(\frac{ N_i}{2}-\epsilon) \frac{\partial F_0}{\partial S_i}\\
W_{eff}^F &=& \sum_{i=0} N_i \frac{\partial F_0}{\partial S_i} + 4 F_1.
\end{eqnarray}
 The first line gives the contribution to the superpotential using the
 unitary embedding theory with adjoint matter whereas the second line
 is the same contribution expressed in the original theory with
 (anti)symmetric matter.  Comparing the two we can extract the
 contribution $F_1$.  We summarize the subleading $1/N$ contributions
 in the different cases in the table
 
 {\large
\begin{center} 
\begin{tabular}{|c|c|}
 \hline
 $U(N)$ + (anti)symmetric  & $F_1 = -\frac{\epsilon}{2}\sum_{i=1}^d \frac{\partial F_0}{\partial S_i}$\\
 \hline
 $SO(N)/SP(N)$ + adjoint  & $F_1 = -\frac{\epsilon}{2}\frac{\partial F_0}{\partial S_0}$\\ \hline
  $SO(N)/SP(N)$ + (anti)symmetric  & $F_1 = -\frac{\epsilon}{2}\sum_{i=1}^d\frac{\partial F_0}{\partial S_i}$\\
  \hline
 \end{tabular} 
\end{center}}
This extends the results for $F_1$ already obtained in \cite{cachazo},
\cite{ll,ikrsv} to the case of unitary gauge group and two-index
tensor representations. In all the theories under consideration the
subleading contributions can be computed from the the leading $F_0$ as
a variation with respect to gaugino bilinear.  Only the gaugino
condensates in $SO/SP$ factor groups contribute!  In particular this
means that for $SO/SP$ with adjoint only the cut at the origin gives
rise to $F_1$, for $SO/SP$ with (anti)symmetric matter all the cuts
contribute and for $U(N)$ with antisymmetric matter only the cut at
the origin does not contribute to $F_1$.


\begin{thebibliography}{77}
\bibitem{DV}{R.~Dijkgraaf and C.~Vafa, ``Matrix models, topological
    strings, and supersymmetric gauge theories,'' Nucl.\ Phys.\ B {\bf
      644}, 3 (2002) [arXiv:hep-th/0206255];\\
%%CITATION = HEP-TH 0206255;%%
    R.~Dijkgraaf and C.~Vafa, ``On geometry and matrix models,''
    Nucl.\ Phys.\ B {\bf 644}, 21 (2002)
    [arXiv:hep-th/0207106];\\
%%CITATION = HEP-TH 0207106;%%
    R.~Dijkgraaf and C.~Vafa, ``A perturbative window into
    non-perturbative physics,'' arXiv:hep-th/0208048.}
%%CITATION = HEP-TH 0208048;%%
\bibitem{cdsw}{F.~Cachazo, M.~R.~Douglas, N.~Seiberg and E.~Witten,
    ``Chiral rings and anomalies in supersymmetric gauge theory,''
    JHEP {\bf 0212}, 071 (2002) [arXiv:hep-th/0211170].}
%%CITATION = HEP-TH 0211170;%%
\bibitem{Gorsky}
A.~Gorsky,
``Konishi anomaly and N = 1 effective superpotentials from matrix models,''
Phys.\ Lett.\ B {\bf 554} (2003) 185
[arXiv:hep-th/0210281].
%%CITATION = HEP-TH 0210281;%%
\bibitem{alday}
L.~F.~Alday and M.~Cirafici,
``Effective superpotentials via Konishi anomaly,''
JHEP {\bf 0305} (2003) 041
[arXiv:hep-th/0304119].
%%CITATION = HEP-TH 0304119;%%
\bibitem{kraus} P.~Kraus and M.~Shigemori, ``On the matter of the
  Dijkgraaf-Vafa conjecture,'' JHEP {\bf 0304} (2003) 052
  [arXiv:hep-th/0303104].
%%CITATION = HEP-TH 0303104;%%
\bibitem{cachazo} F.~Cachazo, ``Notes on supersymmetric Sp(N) theories
  with an antisymmetric tensor,'' arXiv:hep-th/0307063.
%%CITATION = HEP-TH 0307063;%%
\bibitem{ll} K.~Landsteiner and C.~I.~Lazaroiu, ``On Sp(0) factors and
  orientifolds,'' Phys.\ Lett.\ B {\bf 588} (2004) 210
  [arXiv:hep-th/0310111].
%%CITATION = HEP-TH 0310111;%%
\bibitem{ikrsv} K.~Intriligator, P.~Kraus, A.~V.~Ryzhov, M.~Shigemori
  and C.~Vafa, ``On low rank classical groups in string theory, gauge
  theory and matrix models,'' Nucl.\ Phys.\ B {\bf 682} (2004) 45
  [arXiv:hep-th/0311181].
%%CITATION = HEP-TH 0311181;%%
\bibitem{adi} A.~Armoni, A.~Gorsky and M.~Shifman, ``An exact relation
  for N = 1 orientifold field theories with arbitrary
  superpotential,'' Nucl.\ Phys.\ B {\bf 702} (2004) 37
  [arXiv:hep-th/0404247].
%%CITATION = HEP-TH 0404247;%%
\bibitem{land} K.~Landsteiner, ``Konishi anomalies and curves without
  adjoints,'' Nucl.\ Phys.\ B {\bf 700} (2004) 140
  [arXiv:hep-th/0406220].
%%CITATION = HEP-TH 0406220;%%
\bibitem{argurio} R.~Argurio, ``Equivalence of effective
  superpotentials,'' Phys.\ Rev.\ D {\bf 70} (2004) 055012
  [arXiv:hep-th/0405250],
%%CITATION = HEP-TH 0405250;%%
  ``Effective superpotential for U(N) with antisymmetric matter,''
  JHEP {\bf 0409} (2004) 031 [arXiv:hep-th/0406253].
%%CITATION = HEP-TH 0406253;%%
\bibitem{CPSK} T.~E.~Clark, O.~Piguet and K.~Sibold, ``Supercurrents,
  Renormalization And Anomalies,'' Nucl.\ Phys.\ B {\bf 143} (1978)
  445;
%%CITATION = NUPHA,B143,445;%%
  ``The Absence Of Radiative Corrections To The Axial Current Anomaly
  In Supersymmetric QED,'' Nucl.\ Phys.\ B {\bf 159} (1979) 1;
%%CITATION = NUPHA,B159,1;%%
  ``The Gauge Invariance Of The Supercurrent In Supersymmetric QED,''
  Nucl.\ Phys.\ B {\bf 169} (1980) 77.
%%CITATION = NUPHA,B169,77;%%
  K.~Konishi, ``Anomalous Supersymmetry Transformation Of Some
  Composite Operators In Sqcd,'' Phys.\ Lett.\ B {\bf 135} (1984) 439.
%%CITATION = PHLTA,B135,439;%%
  K.~Konishi and K.~Shizuya, ``Functional Integral Approach To Chiral
  Anomalies In Supersymmetric Gauge Theories,'' Nuovo Cim.\ A {\bf 90}
  (1985) 111.
%%CITATION = NUCIA,A90,111;%%
\bibitem{svrcek}
P.~Svrcek,
``On nonperturbative exactness of Konishi anomaly and the Dijkgraaf-Vafa
conjecture,''
JHEP {\bf 0410} (2004) 028
[arXiv:hep-th/0311238].
%%CITATION = HEP-TH 0311238;%%



\end{thebibliography}
\end{document}